\documentstyle[prb,twocolumn,aps]{revtex}

\begin{document}
\tolerance 50000
\title{
Field theory of the fractional quantum Hall effect}
\author{R.Shankar$^{1}$,  and
 Ganpathy Murthy$^{2}$
}
\address{
$^{1}$Sloane Physics Laboratories, Yale University, New Haven CT   
06520 \\
$^{2}$ Department of Physics and Astronomy, Johns Hopkins University,  
Baltimore
MD 21218  \\ and Department of Physics, Boston University, Boston MA  
02215}

\twocolumn[

\date{\today}
\maketitle
\widetext

\vspace*{-1.0truecm}

\begin{abstract}
\begin{center}
\parbox{14cm}{
We present a Chern-Simons theory of the fractional quantum Hall  
effect in which
flux attachment is followed by a transformation that
effectively attaches the correlation holes. We extract the
correlated wavefunctions, compute the drift and cyclotron currents  
(due to
inhomogeneous density), exhibit the Read operator, and operators
that create quasi-particles  and holes.  We show how the bare kinetic  
energy
can get  quenched and replaced by one due to interactions. We find  
that for
$\nu =1/2$ the low energy theory has neutral quasiparticles and give  
the
effective hamiltonian and constraints. }
\end{center}
\end{abstract}

\pacs{
\hspace{1.9cm}
PACS numbers: }

]
\narrowtext
The experimental discovery of the Fractional Quantum Hall
Effect\cite{Tsui,Willet} led to theoretical response on two fronts:  
trial
wavefunctions that captured the essential physics and approximate
computational schemes  starting with the microscopic hamiltonian. The  
most
successful among the latter has been the Chern-Simons (CS) field  
theory
\cite{Gir,Sarma,L,BH,FDM,GM,read1,J,Fetter,LF,ZHK,KZ,HLR,Kwon}. Here  
we present
a formulation of the CS theory which resolves several nagging  
questions and
exposes the
physics in a particularly transparent way. We illustrate our method  
through the
cases $\nu = 1/3$ and $\nu =1/2 $, where the filling fraction  $\nu =  
2\pi
n/eB$, $n$ being the particle density, $-e$ the electron charge and  
$B$ the
magnetic field down the z-axis. Our results for $\nu =
p/(2np+1)$ will be reported later.    We set  $\hbar = c =volume =1,  
z
= x + iy $, and $l_0 = (eB)^{-1/2}$  the cyclotron length.

In the CS approach one introduces a wavefunction for the CS particles  
in terms
of the electronic one as follows:
 \begin{equation}
\Psi_e  =  \prod_{i<j} {(z_i - z_j) ^{l}\over |z_i
-z_j|^{l}}\Psi_{CS}.\label{phase}
\end{equation}
where $l$ is the number of flux quanta to be attached. The prefactor  
introduces
a gauge field ${\bf a}$  obeying
\begin{equation}
{\nabla \times
{\bf a ({\bf r})}\over 2 \pi l} = \sum_i \delta ({\bf r -r}_i  
).\label{const}
\end{equation}
In second quantized form, the CS action
density is
\begin{equation}
S =    \overline{\psi} i \partial_o \psi + a_0 ({\nabla \times {\bf
a}\over 2 \pi l} - \overline{\psi }\psi )  - { |(-i\nabla +e {\bf A}
+{\bf a})\psi|^2 \over 2m}\label{CSaction}
\end{equation}
where ${\bf A}$ is the external vector potential and $m$ is the bare
mass.   The coulomb interaction will  be added  later. By shifting
${\bf a}$ we can cancel  $e{\bf A}$ upon choosing $l=3$ for $\nu  
=1/3$ and
$l=2$ for $\nu = 1/2$.  Hereafter   ${\bf a } $ and
$\psi^{\dag} \psi = \rho $ (the density) will denote
normal-ordered quantities. Whereas in the quest for wavefunctions, we  
can build
in not
just
the phase, but all of $\prod (z_i -z_j)^l$,  or  even the  ubiquitous  
gaussian
factors into the process of flux attachment,  doing so here would  
lead to a
complex vector potential.\cite{RS}. The correlation zeros must  
therefore be
extracted  out of the fluctuations\cite{ZHK}.

We now introduce our scheme (inspired by the work of Bohm-Pines  
\cite{BP}) and
define a composite particle (CP) field, where $P$ may stand for  
fermion $F$ or
boson  $B$:
\begin{equation}
\psi_{CS}(x,y,t) = \exp \left[ -i\int_{-\infty}^{t} a_0(x,y,t') dt'  
\right]
\psi_{CP}(x,y,t)\label{BPF2}.
\end{equation}
The  transformation kills the $a_0\overline{\psi }\psi$ term and
introduces a  longitudinal vector potential $2 \pi l\  {\bf P }$
defined by
\begin{equation}
{\bf q} a_0({\bf q},\omega ) = - \omega 2\pi l{\bf P}({\bf q},\omega
)\label{trans}
\end{equation}
in  the kinetic energy  term, while  $a_0 ({\nabla \times {\bf  
a}\over 2 \pi l
} )$  becomes
\begin{equation}
\sum_q \int d\omega P (-\omega , -q) [-i\omega ] a (q,\omega )
\end{equation}
where $P = -i\hat{{\bf q}}\cdot {\bf P}$ and $a = i \hat{{\bf  
q}}\times {\bf
a}$, {\em so  that the longitudinal and transverse vector potential  
are
now canonically conjugate  and the constraint field has become
dynamical!} The hamiltonian density is
\begin{eqnarray}
H &=& {1 \over 2m}|(-i\nabla  + \ {\bf a} + 2\pi l{\bf P})\psi |^2
\label{hamo}\\
& =&  {1 \over 2m} |\nabla \psi|^2 +{n \over 2m}(a^2+ 4\pi^2l^2 P^2)
\nonumber \\
& +&  ({\bf a} +2\pi l{\bf P}) \cdot {1 \over 2m} \psi^{\dag}(-i
\stackrel{\leftrightarrow}{\nabla} )\psi
 + { 1\over 2m}
:\psi^{\dag}\psi :({\bf a} +2 \pi l {\bf { P} })^2 \nonumber \\
&\equiv &H_0 + H_I + H_{II} \label{ham}
\end{eqnarray}
where $H_I$ and $H_{II}$ refer to the last two terms. Note  that  
$\psi$ is
to be quantized as a boson (fermion) for $\nu = 1/3$ ($1 /2$).
{\em Though $H_I$ and $H_{II}$ denote  interactions between the  
particles  and
the gauge bosons, we are still discussing {\sf free electrons}. We  
are however
 paving
the way for the coulomb interaction.}

The constraint  now defines physical states:

\begin{equation}
({\nabla \times {\bf a}\over 2 \pi l} - :{\psi^{\dag} }\psi :)|  
physical \
\rangle =0.\label{conststates}
\end{equation}
ensuring they are singlets under the local gauge symmetry possessed  
by $H$ and
generated by the operator which annihilates physical states above.
 This constraint is to be expected since we cannot  simply add  extra  
degrees
of
freedom.  From  $H_0$ we see that  the pair $(a,P)$ describe
oscillators at  $\omega_0 = eB/m$. Since  we started
with  $n$ electrons in a plane, $n_0$, the number of independent   
oscillators
obeys  $n_0 \le 2n$.
 {\em We  find that the value
 $n_0 =n$, i.e. for $0<q\le Q = k_F$ recommends itself for many  
reasons and
choose it.}
To pay for these degrees of freedom, the particles will be  deprived  
of $n$
coordinates, which
will be seen to  put  them in the LLL.
 Thus in  Eqn.(\ref{hamo})  only the vector potential ${\bf a}(q)$  
with
$0<q<Q$,   will have a conjugate momentum ${\bf P(q)} $.  For $q>Q$,  
the
short range part  $\delta {\bf a} ({\bf q})$   will    be a function  
of $\rho
(q)$ as in Eqn.(\ref{CSaction}).   The contribution of $\delta {\bf  
a} ({\bf
q})$,  $H_{\delta a}$,  suppressed in Eqn.(\ref{hamo}),  will   be  
discussed
later.

We now analyze $H$ first  ignoring  all but $H_0$, starting
with the case $\nu =1/3$. In the ground state, the bosons condense  
into a
constant wavefunction while the oscillators with hamiltonian
\begin{equation}
H_{osc} =\sum_{q}^{Q} (A^{\dag}(q)A(q))  \omega_0
\end{equation}
where $
  A(q) = (a(q) + 6 \pi  i P(q))/\sqrt{12\pi} $,  yield the ground  
state
wavefunction:
\begin{eqnarray}
\Psi_{} &=& \exp \left[ -\sum_q { 1\over 12\pi}a^2 ({\bf q})  
\right]\\
&=& \exp \left[ -\sum_q { 3\pi}:\rho    ({\bf q}):{1 \over q^2}:\rho     
(-{\bf
q}):  \right]\label{3pi}\\
&=& \prod_{i<j}|z_i -z_j|^3 \exp \left[ -\sum_j  
|z_j|^2/4l_{0}^{2}\right]
\label{corr}
\end{eqnarray}
upon using  the constraint to get the wavefunction in terms of  
particle
coordinates.   The steps leading to the last line are explained in  
Kane
 {\em
et al}\cite{Kane}  and Zhang's review\cite{ZHK}. Putting back the  
phase factors
from  Eqn.(\ref{phase}),
gives us Laughlin's $\psi_{1/3}$ with the proviso that since $q<Q$ in
Eqn.(\ref{3pi}), our answer is to be trusted only for $|z_i -z_j| >  
l_0$:    as
$z_i \to z_j$, we  know that there are three zeros in a
circle of size $l_0$, but not that they coincide.

Let us  understand how not just the phase, but the cubic
correlation zeros of $\psi_{1/3}$,  got built in. Writing  
Eqn.(\ref{BPF2}) in
operator form (at
the origin) \cite{op} as
\begin{equation}
\psi^{\dag}_{CB} = \exp \left[  \sum_{\bf q} {6\pi i \over q} P({\bf  
q})
\right]\psi^{\dag}_{CS}
\end{equation}
we see that when we create a composite boson, we not only create a CS  
boson but
also displace $a(q)$ by $-6\pi /q$, which, {\em upon projection to  
the physical
states} leads to a hole of  charge $-1$ (in electronic units). This  
agrees
with Read's picture \cite{read} of how to add an extra electron to
the $\nu =1/3$ state: we add three units of flux, create a  
correlation hole of
charge  $-1$, and drop in  the newcomer. Evidently $\psi_{CB}^{\dag}$  
is the
Read operator that will have  a nonzero expectation
value in the ground state. Finally, consider
\begin{equation}
\psi_{qh} = \exp \left[  \sum_{\bf q} {2\pi i \over q} P({\bf q})
\right].
\end{equation}
which clearly  creates a Laughlin quasihole of charge $-1/3$. (It  
produces the
extra factor $\prod_i |z_i|$., while the phase   comes  from the  
vortex in the
boson wavefunction.) The adjoint operator creates a quasielectron.   
Our
procedure, which  gives a concrete realization of
many ideas pertaining to composite fermions and bosons was possible  
only upon
 going to a larger space, where collective charge motion is
represented by $a$, with a conjugate momentum $P$ that can be used to  
shift it.

We now turn exclusively to
 $\nu=1/2$ , which was studied exhaustively by Halperin, Lee and Read
(HLR)\cite{HLR}.  First, an   similar analysis  to the one above  
yields the
Rezayi-Read\cite{RR} or Jain wavefunction (quadratic zeros times the  
gaussian,
times a Fermi sea) but without any projection to the LLL.   The  
projection will
be achieved shortly.

Let us first perform two simple calculations.
Imagine adding a smooth weak scalar potential $V(x,y)$. This couples  
via a term
(upon using the constraint)
\begin{equation}
H_V = \int d^2x \  V \  \nabla \times {{\bf a}\over 4 \pi }=-\int  
d^2x  {{\bf
a}\over
4 \pi }\cdot \hat{\bf z} \times \nabla V .
\end{equation}

This linear coupling in ${\bf a}$  shifts ${\bf a}$ to a new minimum  
and leads
to a ground state current
\begin{equation}
<{\bf j}> =  -{e\over 4\pi } \hat{\bf z}
\times \nabla V.
\end{equation}
which implies a  Hall conductance
 (recall $ h/2\pi =1$)
\begin{equation}
\sigma_{xy} = {e^2\over 4\pi  }={e^2 \over 2h}.
\end{equation}
Although the shifted oscillators are in their ground states, the  
original ones
are in an admixture with excited states, as is essential to get the  
right
response\cite{RS}.

Next we confirm that any inhomogeneous density $n(x,y)$  leads to an
uncanceled cyclotron current \cite{SH}
\begin{equation}
<{\bf j}_{cyclo}> =  {e \over 2m} \hat{\bf z} \times \nabla n  
.\label{cyclo}
\end{equation}
To this end imagine  that there is a spatially varying field $B(x,y)$  
and that
a
suitable  scalar potential has been added on top of it to ensure that  
we are
locally at $\nu = 1/2$.  {\em We are thus not calculating any  
standard response
function; our limited goal is to show that this varying density  
$n(x,y)$ leads
to the expected cyclotron current.  } Consider  the last term
$H_{II}$ in Eqn.(\ref{ham})
\begin{eqnarray}
H_{II} &=& {1\over 2m} \int d^2x :\psi^{\dag}\psi : ({\bf a }+ 4\pi   
{\bf
P})^2\nonumber \\
&=& {1\over 2m} \int d^2x :\psi^{\dag}\psi : \langle({\bf a }+ 4\pi   
{\bf
P})^2\rangle + { fluctuations} \nonumber \\
&=& {1\over 2m} \int d^2x \left[ {1 \over 4\pi   } \nabla \times {\bf  
a}
\right]
\ \  4\pi  \  n(x,y) + ...
\end{eqnarray}
where we have used the fact that the zero-point energy density of the  
$n_0 =n$
oscillators implies
\begin{equation}
<({\bf a }+ 4\pi {\bf P})^2> =4\pi \  n(x,y).
\end{equation}
If we shift the oscillators to the new minimum mandated by this  
linear term in
${\bf a}$, we find an average  current given by Eqn.(\ref{cyclo}).

Although answers that only depended on the oscillators  were  
correctly given
above in what we call the middle representation,   the
large kinetic energy of the particles, of order $1/m$,  needs to be
quenched. This will now be done   by  eliminating  the coupling  
between the
fermions and oscillators    by a
canonical transformation that takes us  to the final representation.  
We do this
approximately  by organizing the calculation in powers of
$q$ and keeping just the leading terms at each stage; as well as by   
setting
$\sum_i e^{i(k-q)r_i} = n \delta (k-q)$  and dropping the fluctuating  
part when
the density appears in a product with other operators. The full  
nature of this
approximation is unclear to us, especially when $Q = k_F $ and not  
particularly
small.  In any event, the results, which have many nice features,    
are
good only for long distances.  We drop $H_{II}$ right away since  
$:\psi^{\dag}
\psi :$ is explicitly of order $q$ due to the constraint and  
eliminate $H_I$.
The operators (in first
quantization) transform as follows (upon dropping vector signs when  
obvious and
defining $V_{\pm} = V_x \pm iV_y$ ) .

\begin{eqnarray}
\Omega^{old}&=& e^{-iS} \Omega e^{iS} \\
iS &= &{\sqrt{2\pi} \over m \omega_0}\left[ \sum_q\sum_i\hat{q}_{+}  
p_{i-}
e^{iqr_i}A(q) -
h. c\right]
\\
A^{old}(q)&=& A(q) - {\sqrt{2\pi} \over m \omega_0}\sum_i  
\hat{q}_-p_{i+}
e^{-iqr _i}\label{A}\\
\rho^{old}(q) &=& {q \over \sqrt{8\pi}}(A(q) + A^{\dag}(-q)) -  
il_{0}^{2}\sum_i
(q \times p_i ) \ e^{-iqr_i}\label{rho}\\
0 &=& \sum_i e^{-iqr_i} + {il_{0}^{2}\over 2} \sum_i (q \times p_i) \
e^{-iqr_i} \ (const) \label{cons}\\
H &= &\sum_i {p_{i}^{2} \over 2m} + \sum_{q=0}^{Q}\omega_0  
A^{\dag}(q)A(q)
\nonumber \\
 & -& {1 \over 2mn} \sum_{q=0}^{Q} \sum_i\sum_j p_i \cdot p_j \   
e^{-iq (r_i
-r_j)} +  H_{\delta  a}\label{Hfin}
\end{eqnarray}
(i) The constraint Eqn.(\ref{cons}) does not involve the oscillators.  
As a
result the fermions and oscillators are truly decoupled in this  
leading
approximation and the former face $n$
constraints. These
 and the hamiltonian will commute among themselves in an exact  
canonical
transformation since they did before and will commute to leading   
order in our
approximation. The transformations generated by the constraints  
represent the
gauge transformation of the middle representation, except now the  
particle
hamiltonian must itself be invariant under it. In particular, if  we  
look at
the first   operator in the constraint (the new density) we see that  
it
generates a shift in all the momenta in the limit $q \to 0 $.  We see  
$H$ in
Eqn. (\ref{Hfin}), is invariant if we once again invoke $\sum_i  
e^{i(k-q)r_i} =
n \delta (k-q)$.  This "drifting Fermi" sea ( seen by Haldane in his  
numerical
work) is part of the larger gauge symmetry of the particle  
hamiltonian.
   (ii) If we restrict ourselves to the ground state
of the  oscillators in Eqn.(\ref{rho}) we find $\rho^{old} =  
\bar{\rho}$ (the
second term) which obeys the commutators of magnetic translations in  
the limit
of small $q$:
\begin{equation}
\left[ \bar{\rho }(q), \bar{\rho} (q')\right] = i l_{0}^{2}(q \times  
q') \
\bar{\rho}(q+q')
\end{equation}
an algebra that was studied in detail by Girvin {\em et  
al}\cite{GMP}. Thus we
are able to put  the electrons in the LLL within a standard field
theory by putting  our oscillators in their ground states. {\em  
Notice  that
the fermions  are now dipolar with respect to electronic charge, a  
feature that
has been anticipated by Read\cite{read}  and Jain\cite{Sarma}.}
(iii) The third term in Eqn.(\ref{Hfin}) comes from transforming the  
oscillator
part of $H_0$.  The last, $H_{\delta a}$  stands for  the $\delta  
{\bf a}$ and
$
(\delta {\bf a})^2$ terms of the short-range gauge field.     (iv)  
The $i=j$
terms in the third sum  renormalize $1/m$ downwards, as we decouple  
the
oscillators. We get  $1/m^* =0$ upon using $\sum_q =n$. If we use a  
smaller
$Q$, there  will be a reduction of $1/m*$ to a fraction of $1/m$, (a  
step in
the right direction) but not a full elimination of $m$ dependence in  
the low
energy sector.      The choices $Q>k_F$ lead to a negative effective  
mass and
are not viable.
(v)
The $i\ne j$ terms
summed from $0$ to $Q$ can be traded for minus the sum from $Q$ to  
$\infty$
since they differ by a delta function $\delta (r_i -r_j)$  that  
vanishes on
fermion (and also hard-core boson) wavefunctions.  Let us  combine  
this term
with $H_{\delta a}$.  These large $q$ variables couple to the fermion  
whose
$1/m^* \to
0$ for the choice $n_0 =n$.  Integrating out the fermions in RPA,  we  
find that
this sector  gives
the magnetoplasmon
with the right position and residue. There is no other structure   
since  $x =
\omega
/(qv^*) \to \infty$ since $v^* \to 0$. (vi) While   $1/m =0$ and the  
correct
cyclotron pole and  residue depend on $Q=k_F$ or $n_0 =n$, the  
dipolar nature
of charge, the constraints and the form of the hamiltonian in  
Eqn.(\ref{Hfin})
will be valid even if $Q$ is given a smaller value.

 Having made  the field theory correctly
reproduce the quenched  fermions and dispersionless magnetoplasmon of  
the
noninteracting problem and we are ready to turn on
interactions. We illustrate the procedure with a coulomb interaction  
that is
cut-off at $q= Q$ so that we can treat it entirely
in terms of our oscillators.       In the final representation of $H$
this adds a term (recall Eqn.(\ref{rho}))

\begin{equation}
H_{coul} = {  l_{0}^{4}\over 2 } \sum_{i\ j} \sum_{q=0}^{  Q} {2\pi  
e^2
\over q} ({q}\times p_i )({q}\times p_j) e^{-iq(x_i -x_j)}
\end{equation}
in addition to a term that renormalizes the oscillator frequency to  
$\omega_0 +
 e^2q/4$ and a feeble derivative coupling between the fermions and  
the
oscillator which makes no difference to this order in $e^2$ or $q$.
The $i=j$ term in $H_{coul}$ (which describes the interaction between  
the
electron and the correlation hole
when they separate) now leads to
\begin{equation}
{1 \over m^*} = {e^2l_0 \over 6} \equiv C \ e^2 l_0.
\end{equation}

Note that  $1/m^*$  is not a small $q$ quantity. Our result is just  
an
estimate;  loop diagrams will surely give it   finite, momentum  
dependent
corrections.   It is encouraging that  numerical work \cite{ambmorf}  
(using the
 full  coulomb
interaction) gives  a not too different value of  $C \simeq .2$ .

The  hamiltonian and constraint are (dropping $H_{\delta a})$:
\begin{eqnarray}
H &= &\! \sum_i {p_{i}^{2} \over 2m^*} + \! {  l_{0}^{4} \over  
2}\sum_{i, j\ne
i } \sum_{q}^{ Q} {2\pi e^2 \over q} ({q} \! \times \!p_i)({q}\!  
\times \! p_j)
e^{-iq(r_i -r_j)}\\
0 &=& \sum_i e^{-iqr_i} + {il_{0}^{2} \over 2} \sum_i (q \times p_i)
e^{-iqr_i}\! \! \!
\end{eqnarray}

The constraint  states to this order in $q$ that the density formed  
out
of a putative cyclotron coordinate vanishes, having been spoken for  
by the
oscillators.

One must solve the   above  theory in a way that respects the  
constraints,
i.e., in  a conserving approximation. This has not been done yet to  
our
satisfaction. Given the dipolar nature of charge (the explicit factor  
of $q$ in
$\rho^{old}$) , one may expect that the  $ \rho^{old}-\rho^{old}$  
structure
factor (and its moments) will have two extra powers of $q$ relative  
to a Fermi
liquid. However  the "drifting sea"  might lead to soft modes and  
compensating
inverse powers of q
in the low-frequency response function, so that the static  
compressiblity
remain finite in the limit $q \to  0$ for short range interactions or  
vanishes
as $q$ for Coulomb interactions.

For  $\nu = 1/3$ a similar analysis of the mass  renormalization will  
hold,
but the RPA will proceed very differently because the constraint  
boson  mixes
with the condensate even  at tree level and suppresses low energy  
excitations.

We have presented a CS theory in which the composite particles carry   
flux and
the correlation holes thanks to the additional transformation that  
made the CS
field into dynamical oscillators,
which were then  frozen.  Depriving  the particles of $n$ degrees of  
freedom
led to LLL behavior.
We  derived  the correlated wave functions, drift and cyclotron
currents, explicit operators for creating the quasi-hole and  
quasi-particle,
Read's operator,  and properly traded the bare mass for an effective  
mass based
on
interactions. For $\nu=1/2$ we exhibited the dipolar couplings  
between the
final
quasiparticles,  and derived an effective hamiltonian and  
constraints.
We thank S. Girvin,  S. Kivelson, D.H. Lee, E. Fradkin, F.D.M.  
Haldane,
 J.  Jain, Y.B. Kim, P.A. Lee, A.H. MacDonald, A. Millis,
G.Moore,
V.Pasquier,
D.Pines, N.Read, S.Sachdev, S.Simon, S.Sondhi, A.Stern and Z.  
Tesanovic and
especially B. Halperin for
discussions and  the National Science Foundation for
Grants No. PHY94-07194, DMR 9120525
(RS) and   9311949  (GM).


\begin{references}


\bibitem{Tsui} D.Tsui, H.Stromer and A.Gossard, Phys. Rev. Lett.,  
{\bf 48},
1599, (1982).
\bibitem{Willet} R.L.Willett {\em et al}, Phys. Rev. Lett., {\bf 71},  
3846,
(1993),
Phys. Rev. {\bf B46}, 7344, (1993).

 \bibitem{Gir} For a review see {\em The Quantum Hall Effect}, Edited  
by
R.E.Prange and S.M
Girvin, Springer-Verlag, 1987.
\bibitem{Sarma} For the latest review see {\it Perspectives in  
Quantum Hall
Effects}, Edited by Sankar
Das Sarma and
Aron Pinczuk ( Wiley, New York, 1997). See in particular the articles  
by
Halperin and Jain.
\bibitem
{L}R.Laughlin, Phys. Rev. Lett, {\bf  50}, 1395, (1983).
\bibitem{BH} B.I. Halperin, Helv. Phys. Acta {\bf 56}, 75, (1983).
\bibitem{FDM} F.D.M.Haldane Phys. Rev. Lett.{\bf 51}, 605, (1983).
\bibitem{GM} S.M. Grivin and A.H. MacDonald, Phys. Rev. Lett. {\bf  
58}, 1252,
(1987). See also S. M. Girvin in Ref. (3).
\bibitem{read1}N. Read, Phys. Rev. Let., {\bf 62}, 86, (1989).
\bibitem{J} J.Jain, Phys. Rev. Lett., {\bf 63}, 199, (1989).
\bibitem{Fetter} A.L.Fetter, C.B. Hanna, and R.L. Laughlin, Phys.  
Rev. {\bf
B39}, 9679, (1989), {\em ibid} {\bf 43}, 309, (1991).
\bibitem{LF} A. Lopez and E.Fradkin, Phys. Rev. {\bf B 44}, 5246,  
(1991), {\em
ibid} {\bf 47}, 7080, (1993), Phys. Rev. Lett., {\bf 69}, 2126,  
(1992).

\bibitem{ZHK}S.C. Zhang, H.Hansson and S.Kivelson, Phys. Rev. Lett.,  
{\bf 62},
82, (1989). See the review by S.C. Zhang, in Int. J. Mod. Phys.,
{\bf B6}, 25, (1992).
\bibitem{KZ} V.Kalmeyer and S.Zhang, Phys. Rev.{\bf B46}, 9889,  
(1992).
\bibitem{HLR} B. I. Halperin, P.A.Lee and N.Read, Phys. Rev. {\bf  
B47}, 7312,
(1993).
\bibitem{Kwon}H.J. Kwon, J.B.Marston and A. Houghton, Phys. Rev.  
Let.{\bf 73},
284, (1994) and Phys. Rev. {\bf B52}, 8002, (1995).
\bibitem{RS} R.Rajaraman and S.Sondhi, Int. J. Mod. Phys. B {\bf 10},  
793
(1996).
\bibitem{BP} D. Bohm and D. Pines, Phys. Rev. {\bf 92}, 609,
(1953).\bibitem{Kane} C. L. Kane, S. Kivelson, D.H. Lee and  
S.C.Zhang, Phys.
Rev.
{\bf B 43 }, 3255 (1991).
\bibitem{op} It is possible to go from $\psi_{CS}$ to $\psi_{CF}$ in  
purely
operator language  {\em in the
enlarged  Hilbert space}.
\bibitem{read} N.Read Semi. Cond. Sci. Tech., {\bf 9}, 1859, (1994).
\bibitem{RR} E. Rezayi and N.Read, Phys. Rev. Lett,{\bf 72}, 900,  
(1994).
\bibitem{SH} S.Simon and B.I.Halperin, Phys. Rev. {\bf B48}, 17368,  
(1993),
A.Stern and B. I. Halperin, Phys. Rev. {\bf B52}, 5890, (1995),
\bibitem{GMP} S.M.Girvin, A.H. MacDonald and P. Platzman, Phys. Rev.  
{\bf B33},
2481, (1986).
\bibitem{ambmorf} R. Morf and  d'Ambrumenil Phys.Rev.Lett., {\bf 74},  
5116,
(1995).
\bibitem{Pasq} The dipoles also  arise naturally in the algebraic
approach being deveoped by Haldane and Pasquier. (Private  
communication.)


\end{references}
\end{document}